\input epsf.tex
 \documentstyle[11pt]{article}

\newcommand{\bmat}{\left(\begin{array}}
\newcommand{\emat}{\end{array}\right)}
\def\NPB{Nucl. Phys. B}

\def\yzero{\smash{\hbox{$y\kern-4pt\raise1pt\hbox{${}^\circ$}$}}}

\def\b{\beta}

\def\-{\hphantom{-}}
\def\ov{\overline}
\def\s2{\frac{1}{\sqrt2}}

\def\beq{\begin{equation}}
\def\eeq{\end{equation}}
\def\beqa{\begin{eqnarray}}
\def\eeqa{\end{eqnarray}}
\def\ba{\begin{array}}
\def\ea{\end{array}}

\def\IF{\relax{\rm I\kern-.18em F}}
\def\II{\relax{\rm I\kern-.18em I}}
\def\IP{\relax{\rm I\kern-.18em P}}
\def\IC{\relax\hbox{\kern.25em$\inbar\kern-.3em{\rm C}$}}
\def\IR{\relax{\rm I\kern-.18em R}}

\def\cp{{\cal P}}

\def\Dsl{\,\raise.15ex\hbox{/}\mkern-13.5mu D} 
\def\IZ{Z\kern-.4em  Z}

 \def\cp#1{\relax\ifmmode {\IP\kern-2pt{}_{#1}}\else $\IP\kern-2pt{}_{#1}$\=fi}

\catcode`\@=11
\newdimen\@rotdimen
\newbox\@rotbox

\def\@vspec#1{\special{ps:#1}}
\def\@rotstart#1{\@vspec{gsave currentpoint currentpoint translate
   #1 neg exch neg exch translate}}
\def\@rotfinish{\@vspec{currentpoint grestore moveto}}
\def\@rotr#1{\@rotdimen=\ht#1\advance\@rotdimen by\dp#1%
   \hbox to\@rotdimen{\hskip\ht#1\vbox to\wd#1{\@rotstart{90 rotate}%
   \box#1\vss}\hss}\@rotfinish}
\def\@rotl#1{\@rotdimen=\ht#1\advance\@rotdimen by\dp#1%
   \hbox to\@rotdimen{\vbox to\wd#1{\vskip\wd#1\@rotstart{270 rotate}%
   \box#1\vss}\hss}\@rotfinish}%
\def\@rotu#1{\@rotdimen=\ht#1\advance\@rotdimen by\dp#1%
   \hbox to\wd#1{\hskip\wd#1\vbox to\@rotdimen{\vskip\@rotdimen
   \@rotstart{-1 dup scale}\box#1\vss}\hss}\@rotfinish}%
\def\@rotf#1{\hbox to\wd#1{\hskip\wd#1\@rotstart{-1 1 scale}%
   \box#1\hss}\@rotfinish}%
\def\rotate{\@ifnextchar[{\@rotate}{\@rotate[l]}}
\def\@rotate[#1]#2{\setbox\@rotbox=\hbox{#2}\@nameuse{@rot#1}\@rotbox}

\catcode`\@=12

\begin{document}

\makeatletter
\@addtoreset{equation}{section} \makeatother
\renewcommand{\theequation}{\thesection.\arabic{equation}}
\pagestyle{empty}
\pagestyle{empty}
\rightline{\today}
\vspace{0.5cm}
\setcounter{footnote}{0}

\begin{center}
{\LARGE{\bf 4D GUT (and SM)  MODEL BUILDING 
FROM INTERSECTING D-BRANES }}
\\[7mm]
{\Large{{  Christos ~Kokorelis } }
\\[2mm]}
\small{\em 
 Institute of Nuclear Physics, N.C.R.P.S. Demokritos,
GR-15310, Athens, Greece} 
\end{center}
\vspace{3mm}


\begin{center}
{\small \bf ABSTRACT}
\end{center}
\begin{center}
\begin{minipage}[h]{14.5cm}

We provide a general overview of the current state of the art in 
three generation model building proposals - using intersecting 
D-brane toroidal compactifications 
of IIA string theories - which have, only, the SM at low energy. In this 
context,  
we focus on these model building directions, where natural non-supersymmetric 
constructions based on
 $SU(4)_C \times SU(2)_L \times SU(2)_R$,   
SU(5) and flipped SU(5) GUT groups, 
have at low energy only the 
Standard Model. In the flipped SU(5) GUTS, the special build up structure 
of the models accommodates naturally a see-saw mechanism and a new solution to the
doublet-triplet splitting problem.

\end{minipage}                 
\end{center}

\newpage
\setcounter{page}{1}
\pagestyle{plain}
\renewcommand{\thefootnote}{\arabic{footnote}}
\setcounter{footnote}{0}

\section{Introduction to SM D6/D5-brane model building with only
the SM at Low Energy}

In the last two years, constructions based 
on D-branes intersecting at angles -intersecting brane worlds (IBW's) for 
short - received a lot of attention \cite{iba} -  
\cite{kokosnew}, as on these constructions, 
for the first time in string theory, it became 
possible \footnote{See also \cite{rev} for other reviews on the subject} 
to construct four dimensional non-supersymmetric 
intersecting D6-brane 
models that have only the
SM gauge group and chiral spectrum - with right handed neutrinos $\nu_R$'s 
- at low energies \cite{iba, kokos1, kokos2}.
The initial constructions of IBW's exhibiting the SM at low energy,
were based on a background of D6-branes intersecting at angles in
4D toroidal orientifolds \cite{lu} of [T-dual to models with magnetic 
deformations \cite{ang}. See also \cite{pra}.] type
IIA theory;
possess broken supersymmetry on the bulk and on 
the `branes', and exhibit proton stability as baryon number is a gauged 
symmetry.
The primary common phenomenological characteristics of the four stack
D6-models of \cite{iba} and the five and six SM's of \cite{kokos1} and 
\cite{kokos2} respectively - emphasizing that there are 
no D6-brane models with only the SM at low energy using 
constructions with more than six stacks of D6-branes at the 
string scale $M_s$ -
are : \newline
$\bullet$  the prediction of the 
existence of the 
SM chiral spectrum together with $\nu_R$'s, 
\newline
$\bullet$ the 
conservation of lepton number - the models admit Dirac terms for the 
neutrinos -  their masses appear as a result of the 
existence of particular Yukawa couplings associated with the breaking
of the chiral symmetry.\newline
Additionally, the SM's of \cite{kokos1} and 
\cite{kokos2} {\em exhibit a new phenomenon} - not found in the SM's of \cite{iba},
namely {\em the prediction of the existence of 
N=1 supersymmetric (SUSY) 
partners of 
$\nu_R$'s, the s$\nu_R$'s.} 
Thus even though the models of \cite{kokos1, kokos2}
are non-supersymmetric, they have particular N=1 SUSY partners, the s$\nu_R$'s,
whose presence is necessary in order to 
break the extra beyond the hypercharge U(1)'s that 
survived massless the presence of the generalized Green-Schwarz
mechanism, the latter cancelling the mixed U(1)-gauge anomalies \footnote{
We note that in these kind of constructions uncancelled NS 
tadpoles remain, whose cancellation in higher orders of
perturbation theory remains an open issue. The NS tadpoles do not 
affect the low energy spectrum of the models but they rather imply 
the instability of the present models, in the present order of perturbation 
theory, in a flat background. Higher order corrections to NS tadpoles
might be responsible for stabilizing these D-brane configurations. 
}.
Starting from a SM-like configuration at $M_s$ and constructing a D-brane
configuration that had only the SM at low energy, as in the models
of \cite{iba, kokos1, kokos2}, was not the only model building 
success story of IBW's. For dimensional (4D) configurations with 
intersecting 
D5-branes and only the SM at low energy - wrapped on a background 
of type IIB compactifications on a $T^4 \times C/Z_N$ - were also 
obtained in  
\cite{kokoiba}.\newline
Moreover the first non-SUSY constructions of 
string GUTS
which have only the SM at low energy \footnote{
These models were also based
on the intersecting D6-backgrounds of \cite{lu}.}, were
constructed in \cite{kokos5}, based on the Pati-Salam structure 
$SU(4)_C \times SU(2)_L \times SU(2)_R$ at $M_s$.

\section{Building the $SU(4)_C \times SU(2)_L \times SU(2)_R$ GUTS With Only
The SM At Low Energy} 

Extensions of these GUTS with four, five and six stacks of D6-branes were also considered in \cite{kokos5, kokos6, kokos7}. 
\newline
The basic features found in these intersecting 
D6-brane models can be classified as 
follows : \newline 
$\bullet$ The models even though they have overall N=0 SUSY,
possess N=1 SUSY subsectors which are necessary in order to create a 
Majorana mass term for $\nu_R$'s. \newline 
$\bullet$ Extra branes are needed to cancel RR tadpoles. The presence of 
these
branes creates extra matter singlets, transforming under both the visible SM
gauge group and the extra D6-brane gauge group that may be 
used to break the extra U(1)'s, beyond hypercharge, surviving massless the 
presence of the generalized Green-Schwarz mechanism. Their presence is also 
used to make massive the exotic fermions, seen for example in the 
bottom part of table (1), taken from \cite{kokos7}.  
\begin{table}[htb] \footnotesize
\renewcommand{\arraystretch}{0.8}
\begin{center}
\begin{tabular}{|c|c||c|c||c||c|c|c|}
\hline
Fields &Intersection  & $\bullet$ $SU(4)_C \times SU(2)_L \times SU(2)_R$
 $\bullet$&
$Q_a$ & $Q_b$ & $Q_c$ & $Q_d$ & $Q_e$\\
\hline
 $F_L$& $I_{ab^{\ast}}=3$ &
$3 \times (4,  2, 1)$ & $1$ & $1$ & $0$ &$0$ &$0$\\
 ${\bar F}_R$  &$I_{a c}=-3 $ & $3 \times ({\ov 4}, 1, 2)$ &
$-1$ & $0$ & $1$ & $0$ & $0$\\
 $\chi_L^1$& $I_{bd^{\star}} = -8$ &  $8 \times (1, {\ov 2}, 1)$ &
$0$ & $-1$ & $0$ & $-1$ & $0$\\    
 $\chi_R^1$& $I_{cd} = -8$ &  $8 \times (1, 1, {\ov 2})$ &
$0$ & $0$ & $-1$ &$1$ &$0$\\
 $\chi_L^2$& $I_{be} = -4$ &  $4 \times (1, {\ov 2}, 1)$ &
$0$ & $-1$ & $0$ &$0$ & $1$ \\    
 $\chi_R^2$& $I_{ce^{\ast}} = -4$ &  $4 \times (1, 1, {\ov 2})$ &
$0$ & $0$ & $-1$ & $0$ &$-1$ \\\hline
 $\omega_L$& $I_{aa^{\ast}}$ &  $6 \b^2
 \times (6, 1, 1)$ & $2$ & $0$ & $0$ &$0$ &$0$\\
 $y_R$& $I_{aa^{\ast}}$ & $6  \b^2  \times ({\bar 10}, 1, 1)$ &
$-2$ & $0$ & $0$ &$0$ &$0$ \\
\hline
 $s_R^1$ & $I_{dd^{\ast}}$ &  $16 \b^2
 \times (1, 1, 1)$ & $0$ & $0$ & $0$ &$2$ &$0$\\
 $s_R^2$ & $I_{ee^{\ast}}$ & $8  \b^2  \times (1, 1, 1)$ &
$0$ & $0$ & $0$ &$0$ &$-2$ \\
\hline
\end{tabular}
\end{center}
\caption{\small Fermionic spectrum of the $SU(4)_C \times
SU(2)_L \times SU(2)_R$, PS-II class of models together with $U(1)$
charges. We note that at energies of order $M_z$ only
the Standard model survives.
\label{spectrum8}}
\end{table}
The fermion spectrum of table (1) is consistent with the calculation of RR
tadpoles. The RR tadpoles get cancelled with the introduction of extra U(1) 
branes, $h^i$, that transform under the both the extra U(1) gauge group 
and the 
rest of the
intersecting D6-branes of table (1). The existence of N=1 SUSY at 
the intersections $dd^{\star}$, $dh$, $dh^{\star}$, $eh$, $eh^{\star}$,   creates the singlets $s_1^B$, $\kappa_3^b$, $\kappa_4^b$, 
$\kappa_5^b$, $\kappa_6^b$ respectively, that contribute to the mass of the
`light' fermions $\chi_L^1$, $\chi_L^2$. All fermions of table (1) receive 
a mass of order $M_s$; the only exception being the light masses of
$\chi_L^1$, $\chi_L^2$, weak fermion doublets. 
Lets us discuss the latter issue in more detail.
\newline
The left handed fermions $\chi_L^1$ receive a contribution to their mass
from the coupling \footnote{
In (\ref{ka1sa1}) we have included the leading contribution of the 
worksheet area connecting the
seven vertices. In the following for simplicity reasons we will set the 
leading contribution of the different couplings to
one e.g. area tends to zero.}
\beq
(1, 2, 1)(1, 2, 1) e^{-A}
 \frac{\langle h_2 \rangle \langle h_2 \rangle
\langle {\bar F}_R^H  \rangle \langle H_1 \rangle
\langle {\bar s}_B^1 \rangle}{M_s^4}
\stackrel{A \rightarrow 0}{\sim}
\frac{\upsilon^2}{M_s} \ (1, 2, 1)(1, 2, 1)
\label{ka1sa1}
\eeq 
and from another coupling, of the same order as (\ref{ka1sa1}), also
contributing to the mass of the $\chi_L^1$ fermion as 
\begin{eqnarray}
(1, 2, 1)(1, 2, 1) \frac{ \langle h_2 \rangle \langle h_2 \rangle
\langle {\bar F}_R^H \rangle \langle H_1 \rangle
 \langle {\bar \kappa}_3^B \rangle
 \langle {\bar \kappa}_4^B \rangle }{M_S^9} \sim
(1, 2, 1)(1, 2, 1)\frac{\upsilon^2}{M_s},
\label{addi1}
\end{eqnarray}
The left handed fermions $\chi_L^2$
receives a non-zero mass
from the coupling 
\beq
(1, 2, 1)(1, 2, 1) 
 \frac{\langle h_2 \rangle \langle h_2 \rangle
\langle {\bar F}_R^H  \rangle \langle H_1 \rangle
\langle  {\bar s}_B^2 \rangle}{M_s^4}
\stackrel{A \rightarrow 0}{\sim}
\frac{\upsilon^2}{M_s} \ (1, 2, 1)(1, 2, 1)
\label{ka1sa2}
\eeq
and the coupling
\begin{eqnarray}
(1, 2, 1)(1, 2, 1) \frac{ \langle h_2 \rangle \langle h_2 \rangle
\langle {\bar F}_R^H \rangle \langle H_1 \rangle
 \langle {\bar \kappa}_5^B \rangle
 \langle {\bar \kappa}_6^B \rangle }{M_S^5} \sim
(1, 2, 1)(1, 2, 1)\frac{\upsilon^2}{M_s},
\label{addi2}
\end{eqnarray}
Thus assuming that the leading area Yukawa for
the couplings is of order ${\cal O}(1)$, e.g. associated areas going to zero,
the masses of \footnote{In this case the masses of $\chi_L^1$, $\chi_L^2$ are the sum of t
he contributions
of (\ref{ka1sa1}, \ref{addi1}) and (\ref{ka1sa2}, \ref{addi2}) respectively
}
\begin{equation}
\chi_L^1,\  \chi_L^2 \sim \frac{2 \upsilon^2}{M_s} \,
\end{equation}
$\bullet$ As the particles $\chi_L^1,\  \chi_L^2$ are not observed at present,
the fact that their mass may be between 
\beq
100 \ GeV \ \ \leq \ \chi_L^1, \ \chi_L^1 \  \leq \  2 \upsilon = \
max \{ \frac{2\upsilon^2}{M_s}\} =\ 492 \ GeV
\eeq
sends the string scale 
\beq
M_s  \leq 1.2 \ TeV
\eeq
This is a general feature of all the Pati-Salam models based 
on toroidal orientifolds; they 
 predict the existence of light weak doublets with masses \footnote{The reader may convince
 itself that the maximum value of $2 \upsilon^2 / M_s$ is $2 \upsilon$.}  
 between 100 and
 $\upsilon = 492$ GeV.
 The latter result may be considered
 as a general
 prediction of all classes of models based on intersecting
 D6-brane Pati-Salam GUTS.
  \newline 
 Another important property of these constructions is that
the conditions for some intersections to respect N=1 supersymmetry and also
needed to guarantee the existence of a Majorana mass term for s$\nu_R$'s : 
\newline  
$\bullet$ solve the orthogonality conditions for the extra - beyond 
hypercharge - U(1)'s \footnote{The latter becoming massive from the use of 
extra singlets created by the presence of extra branes; the latter
 needed to satisfy the 
RR tadpoles.} to survive massless the presence of a generalized
Green-Schwarz mechanism describing the couplings of the U(1)'s
to the RR two form fields. \newline   
The considerations we have just described \cite{kokos5}, \cite{kokos6}, 
\cite{kokos7} are quite generic and 
the same methodology
applies easily
to the construction of more general GUT gauge groups in the context of
intersecting brane worlds.

We note that at present the only existing string GUT constructions,
in the context of Intersecting D6-brane Models, 
that have only the SM at low energy, with complete cancellation 
of RR tadpoles, are: \newline 
a) the toroidal orientifold II Pati-Salam GUTS of 
\cite{kokos5, kokos6, kokos7} and 
\newline
b) the constructions of flipped SU(5), and 
SU(5) GUTS of \cite{kokos4} described next.

\section{The Construction of Flipped SU(5) (and SU(5)) GUTS with only
the SM at Low Energy} 

Lets us review the intersecting D6-branes constructions of 
the $Z_3$ orientifolds of \cite{lust3}. The D6-branes involved satisfy
the following RR tadpole conditions where 
\footnote{
The net number of bifundamental massless chiral fermions in the models
is defined as
\beqa
({\bar N}_a, N_b)_L :\  I_{ab} = Z_a Y_b - Y_a Z_b \\
(N_a, N_b)_L : \   I_{ab^{\star}} = Z_a Y_b + Y_a Z_b
\label{spec1}
\eeqa
}
\beq
\sum_a N_a Z_a = 2 \ .
\label{tad}
 \eeq
As it was noticed in \cite{lust3} the simplest realization of an SU(5) GUT
involves two stacks of D6-branes at the string scale $M_s$, the
first one corresponding to a $U(5)$ gauge group while the second one to a
U(1) gauge group. Its effective wrapping numbers are given by 
\beq (Y_a,
Z_a) = (3, \frac{1}{2}), \  (Y_b, Z_b) = (3, -\frac{1}{2}),
\label{ena} \eeq Under the decomposition $ U(5) \subset SU(5)
\times U(1)_a$, the models become effectively an $SU(5) \times
U(1)_a \otimes U(1)_b$ GUT. One combination of U(1)'s
become massive due to its coupling to a RR field, 
another one remains massless to low energies. The spectrum of this
SU(5) GUT may be seen in the first seven columns (reading from the left) 
of table (2).
\begin{table}[htb]\footnotesize
\renewcommand{\arraystretch}{1.5}
\begin{center}
\begin{tabular}{||c|c|c|c|c|c|c|c||}
\hline \hline
Field & Sector name & Multiplicity & $SU(5)$    &  $U(1)_a$ & $U(1)_b$
 & $U(1)^{mass}$ & $U(1)^{fl} = \frac{5}{2} \times U(1)^{mass} $\\\hline
$f$& \{ 51 \} & $3$          & ${\bf {\bar 5}}$ & $-1$      &  $1$   & $-\frac{6}{5}$ & -3    \\\hline
$F$ & $A_{a}$ & $3$          & ${\bf 10}$   &  2         &   0      &$\frac{2}{5}$ &   1          \\\hline
$l^c$ & $S_{b}$ & $3$          &  ${\bf 1} $         &   0        &   -2   & $2$ & 5    \\\hline
  \hline
\end{tabular}
\end{center}
\caption{Chiral Spectrum of a two intersecting D6-brane stacks in a three 
generation flipped $SU(5) \otimes U(1)^{mass}$ model.
Note that the charges under the $U(1)^{fl}$ gauge symmetry,
 when rescaled
appropriately (and $U(1)^{fl}$ gets broken) `converts' the flipped SU(5) model
to a three
generation (3G) $SU(5)$. 
\label{flip}}
\end{table}
At this stage the SU(5) models - have the correct chiral 
fermion content of an SU(5) GUT - and the extra U(1) surviving 
the presence of the Green-Schwarz mechanism, 
breaks by the use of a singlet field present. However, 
the electroweak ${\bf }5$-plets needed for electroweak
 symmetry breaking of the models are absent. Later on, attempts to 
construct a
fully N=1 supersymmetric SU(5) models at $M_s$ in \cite{cve4}, produced 
3G models
that were not free of remaining massless exotic 15-plets. 
Also, later on 
in \cite{nano} it was noticed that if one leaves unbroken, 
and rescales, the U(1) 
surviving massless the Green-Schwarz mechanism of the SU(5) GUT of 
\cite{lust3}, the rescaled U(1) becomes
the flipped U(1) generator. However, the proposed 3G models lack the presence
of GUT Higgses or electroweak pentaplets and were accompanied by extra
exotic massless matter to low energies.

In \cite{kokos4} we have shown that it is possible to construct the first 
examples of string 
SU(5) and
 flipped
SU(5) GUTS - where we identified the appropriate GUT and electroweak 
Higgses - which break to the SM at low energy.
E.g. in the flipped SU(5) GUT, the fifteen fermions of the SM plus the right
handed neutrino $\nu^c$ belong to the 
\beq
F = {\bf 10_1} = (u, d, d^c, \nu^c), \ \
f = {\bf {\bar 5}_{-3}} = (u^c, \nu, e), \  \ l^c =  {\bf 1_{5}} = e^c
\label{def1}
\eeq
chiral multiplets.
The GUT breaking Higgses may come from the `massive' spectrum
of the sector localizing
the ${\bf 10}$-plet (${\bf 10_1^B} = (u_H, d_H, d^c_H, \nu^c_H )$ 
fermions seen in table (\ref{flip}). The lowest order
Higgs in this sector, let us call them $H_1$, $H_2$, have quantum numbers as
those given in table (\ref{Higgsfli}).
\begin{table} [htb] \footnotesize
\renewcommand{\arraystretch}{1}
\begin{center}
\begin{tabular}{||c|c||c|c|c|c||}
\hline
\hline
Intersection & GUT Higgses & repr. & $Q_a$ & $Q_b$ & $Q^{fl}$\\
\hline\hline
$\{ a,{\tilde O6} \}$  & $H_1$  &  {\bf 10}   & $2$   & $0$ & $1$ \\
\hline
$ \{ a,{\tilde O6} \}$  & $H_2$  &  ${\bf {\bar 10}}$   & $-2$ & $0$  & $-1$ \\
\hline
\hline
\end{tabular}
\end{center}
\caption{\small 
Flipped $SU(5) \otimes U^{fl}$ GUT symmetry breaking
scalars. 
\label{Higgsfli}}
\end{table}
By looking at the last column of table (\ref{Higgsfli}), we realize that the
Higgs $H_1$, $H_2$ are the GUT symmetry breaking Higgses of a standard
flipped SU(5) GUT.
By dublicating the analysis of section (3.1),
one may conclude that what it appears in the effective theory as GUT 
breaking Higgs scalars, is the combination 
$
H^G = H_1 + H_2^{\star}$.
In a similar way the correct identification
of the electroweak content \cite{kokos4} of the flipped SU(5) ${\bf 5_{-2}^B} = (D, h^{-},  h^{0} )$-plet
(and 
SU(5))GUTS made possible 
the existence of the see-saw mechanism which is generated by the 
interaction \beqa {\cal
L} = {\tilde Y}^{\nu_L \nu_R} \cdot {\bf 10} \cdot {\bar {\bf
5}} \cdot {\bar h}_4 \
      + \  {\tilde Y}^{\nu_R} \cdot \frac{1}{M_s} \cdot ({\bf 10} \cdot
{\bf \overline{10}}^B)
      ({\bf 10} \cdot {\bf \overline{10}}^B)\ . 
\label{seesaw1}
\eeqa
Its standard version can be generated by choosing
\beq
\langle h_4 \rangle = \upsilon, \  \langle {\bf 10}_i^B \rangle = M_s
\label{masse1}
\eeq
and generates small neutrino masses.
In these constructions the baryon number is not a gauged symmetry, thus 
a high GUT scale of the order of the $10^{16}$ GeV helps the theory to
avoid gauge mediated proton decay modes like the \cite{kokos4}
\beq
  \sim \frac{1}{M_s^2}\ ({\bar u}^c_L \ u_L) \ ({\bar e}_{R}^{+}) 
(d_{R}),\ \
\sim \frac{1}{M_s^2}\ ({\bar d}^c_R \ u_R) ({\bar d}^c_L \ \nu_L) \ .
\eeq   
[In IBW's proton decay by direct calculation of string amplitudes
for SUSY SU(5) D-brane models was examined in \cite{igorwi}.]
Scalar mediated proton decay modes get suppressed by the existence of 
a new solution to the doublet-triplet splitting problem   
 \beq
\frac{r}{M_s^3}(HHh)( {\bar F}  {\bar F} {\bar h}) + m ({\bar h}h) (
{\bar H} H) + \kappa ({\bar H}H)( 
{\bar H} H),\eeq
that stabilizes the vev's of the triplet scalars $d_c^H$, $D$ \cite{kokos4}.
This is the first example of a doublet-triplet splitting realization
in IBW's. 
The full solution of the gauge hierarchy problem,
that is avoiding the existence of quadratic corrections to the electroweak 
Higgses remains an open issue in the present GUTS.  

 Recently the interest of model building in IBW's has been focused in 
the 
construction of intersecting D6-brane models which localize the 
spectrum of MSSM at low energies
\cite{ibanew}, \cite{kokosnew}.

\end{document}